\documentclass[aps,prl,twocolumn,superscriptaddress,amsmath,amssymb,showpacs]{revtex4-1}
\usepackage[pass]{geometry}
\usepackage{graphicx,bm,epsfig,amssymb,amsfonts,braket,color}
\usepackage[resetlabels]{multibib}
\newcommand{\ba}{\begin{eqnarray}}
\newcommand{\ea}{\end{eqnarray}}
\newcommand{\bd}{\begin{displaymath}}
\newcommand{\ed}{\end{displaymath}}

\newcommand{\bpm}{\begin{pmatrix}}
\newcommand{\epm}{\end{pmatrix}}
\newcommand{\nn}{\nonumber \\}

\begin{document}
\title{When Chiral Photons Meet Chiral Fermions -- \\Photoinduced Anomalous Hall Effects in Weyl Semimetals}
\author{Ching-Kit Chan}
\affiliation{Department of Physics, Massachusetts Institute of Technology, Cambridge, Massachusetts 02139, USA}
\author{Patrick A. Lee}
\affiliation{Department of Physics, Massachusetts Institute of Technology, Cambridge, Massachusetts 02139, USA}
\author{Kenneth S. Burch}
\affiliation{Department of Physics, Boston College, Chestnut Hill, Massachusetts 02467, USA}
\author{Jung Hoon Han}
\email{hanjh@skku.edu}
\affiliation{Department of Physics, Sungkyunkwan University, Suwon, 440-746, Korea}
\email{hanjh@skku.edu}
\author{Ying Ran}\email{ying.ran@bc.edu}
\affiliation{Department of Physics, Boston College, Chestnut Hill, Massachusetts 02467, USA}
\email{ying.ran@bc.edu}
\date{\today}

\begin{abstract}
The Weyl semimetal is characterized by three-dimensional linear band touching points called Weyl nodes. These nodes come in pairs with opposite chiralities. We show that the coupling of circularly polarized photons with  these chiral electrons generates a Hall conductivity without any applied magnetic field in the plane orthogonal to the light propagation. This phenomenon comes about because with all three Pauli matrices exhausted to form the three-dimensional linear dispersion, the Weyl nodes cannot be gapped. Rather, the net influence of chiral photons is to shift the positions of the Weyl nodes. Interestingly, the momentum shift is tightly correlated with the chirality of the node to  produce a net anomalous Hall signal. Application of our proposal to the recently discovered TaAs family of Weyl semimetals leads to an order-of-magnitude estimate of the photoinduced Hall conductivity which is within the experimentally accessible range.
\end{abstract}
\maketitle

%%%%%%%%%%%%%%%%%%%%%%%%%%%%%%%%%%%%%%%%%%%%%%%%%%%%%%%%%%%%%%%%%%%%%%%%%%%%%%%%%%%%%%%%%%%%%%%%%%%%%%%%%%%%%%%%%%%%%%%%%%%%%%%%%%%%%%%%%%
{\it Introduction} - Previously thought to be an asset exclusive to massless elementary particles such as photons, chirality has now become a defining emergent property of electrons in such crystalline materials like graphene~\cite{RevModPhys.81.109} and the surface of topological insulators~\cite{RevModPhys.82.3045}. An exciting new addition to the growing list of ``chiral electronic materials''is the Weyl semimetal~\cite{Herring,murakami,wan11,Yang:2011p75129,PhysRevLett.107.127205}. In this three-dimensional analogue of graphene, the Weyl nodes act as magnetic monopoles in momentum space~\cite{Fang03102003}. In sharp contrast to the two Dirac nodes in graphene that behave more or less as separate low-energy degrees of freedom, a pair of Weyl nodes carrying opposite chiralities are inherently tied together in a non-local manner through the ``chiral anomaly'' mechanism which allows the dissipationless transfer of charge between them~\cite{Nielsen1983389}. The subject has undergone vigorous research lately due to reports of material predictions, followed immediately by their synthesis and confirmation of their Weyl characters by photoemission experiments and non-local transport measurements. A striking example of success along this line in recent years is the transition metal monophosphide family that includes TaAs, TaP, NbAs, and NbP~\cite{Dai:2015p11029,Huang.S.M:2015,Xu07082015,HongDing,Lee:2015tz,Lv:2015kp,Yang:2015ev,Xu.S.Y:2015,Huang:2015gy,2015arXiv150302630Z}.

\begin{figure}[b]
\begin{center}
\includegraphics[angle=0, width=.6\columnwidth]{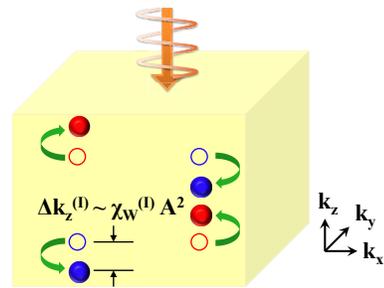}
\caption{(Color online) Schematic figure for a driven Weyl semimetal. Blue and red circles indicate Weyl nodes with opposite chiralities $\chi_W^{(I)}$. The node positions are shifted by the chiral photons in a chirality-dependent manner and the shift is proportional to the square of the driving amplitude $A$. As a result, even though the total momentum shift along the driving direction is zero ($\sum_I \Delta k_z^{(I)}=0$), the overall Chern vector shift has a finite $z$-component ($\delta \nu_z = \sum_I \chi_W^{(I)} \Delta k_z^{(I)} \neq 0$), resulting in a photoinduced anomalous Hall conductivity $\sigma_{xy}$ (see Eq.~(\ref{eq:anomalous_Hall})).}
\label{fig_schematic_setup}
\end{center}
\end{figure}

In light of the rapid maturity of the Weyl semimetal research, one should ask whether the unusual electronic transport properties in both the static and dynamic regimes of the Weyl semimetal can be exploited in high speed electronics or to provide a new means of revealing dynamic topological effects~\cite{Huang:2015gy,2015arXiv150302630Z,2015NatPh..11..645S}. It was recently suggested that exposing a two-dimensional Dirac material to a circularly polarized (CP) light could lead to a novel form of Hall effect due to the effective gap opening at the Dirac point~\cite{OkaAoki,Wang25102013}. A natural question arises as to what happens when the three-dimensional, linearly dispersing band of electrons couple to an intense CP light, as schematically presented in Fig.~\ref{fig_schematic_setup}, and whether a non-trivial Hall effect can be induced. Here we present a direct consequence of coupling CP light to the Weyl fermion: the appearance of the anomalous Hall effect (AHE).

%While previous work by one of the authors (YR)~\cite{Yang:2011p75129} has given a general formula for anomalous Hall conductivity arising from Berry curvature effect in the Weyl semimetal, but no effort had been made at that time to tie it to realistic materials.

In this work, we show that even when the AHE is absent due to symmetry reasons in typical Weyl materials, it must be generally present in all Weyl systems when coupled to a CP light source. Based on a low-energy effective Weyl-Floquet Hamiltonian analysis and symmetry considerations, we discuss the induced AHE as a generic and readily observable phenomenon in Weyl semimetals. We further demonstrate this effect using a concrete microscopic model, and apply our study to the TaAs family of Weyl semimetals with an estimation of the experimental feasibility. Other interesting works have addressed the driven three-dimensional Dirac electrons~\cite{PhysRevB.91.205445,oka15} or gapped systems~\cite{Lindner2011,0295-5075-105-1-17004}. However, neither systems enjoy the unique topological protection of the Weyl nodes, whose response, under a CP beam, results in the experimental signature of the AHE as we explain below.
\\

%%%%%%%%%%%%%%%%%%%%%%%%%%%%%%%%%%%%%%%%%%%%%%%%%%%%%%%%%%%%%%%%%%%%%%%%%%%%%%%%%%%%%%%%%%%%%%%%%%%%%%%%%%%%%%%%%%%%%%%%%%%%%%%%%%%%%%%%%%
{\it General tight-binding model analysis}: The low-energy Hamiltonian for a single Weyl node can be parameterized most generally by a real 3-vector $\vec \alpha$ and a $3\times 3$ real matrix $\beta$:
\ba
H_W(\vec q) = q_i\alpha_i \sigma_0+q_i\beta_{ij} \sigma_j, \label{eq:two-band-Weyl}
\ea
with a canonical 3D linear dispersion $E_W(\vec q)=\alpha_i q_i \pm [\sum_j (\sum_i q_i \beta_{ij})^2]^{1/2}$. Einstein summation is assumed in the momentum $q_i$ and the Pauli matrix $\sigma_j$. The chirality of the Weyl node is given by the sign of the determinant of $\beta$: $\chi_W=\mbox{sgn}[\mbox{Det}(\beta)]$. For a practical band structure, this 2$\times$2 Weyl Hamiltonian is embedded within a large $N$-band Hamiltonian as a low-energy sector. In order to have the most general consideration of the photoinduced AHE, we therefore consider a $N$-band tight-binding Hamiltonian $H(\vec k)$. The expansion of $H(\vec k)$ around a Weyl node at $\vec k_{W}$ to first order in $\vec q = \vec k-\vec k_W$ can be put in the block form:
\ba
U^{\dagger} H(\vec k_W + \vec q) U
&=&H_{\rm lin} (\vec q ) +O(q^2), \nn
H_{\rm lin} (\vec q) &=& \begin{pmatrix}
H_{\rm W}(\vec q) & q_i C_{i}\\
q_i C^{\dagger}_{i}& D_0+q_i D_i
\end{pmatrix},
\ea
where the matrix $D_0$ gives the $(N-2)$ high-energy states at $\vec k = \vec k_W$. Eigenvectors at $\vec k= \vec k_W$ are used to construct the unitary matrix $U$ whose first two columns corresponding to the two zero-energy states. In general, the $2\times (N-2)$ matrices $C_i$ mix the Weyl bands with high-energy bands linearly in $\vec q$ and should not be ignored.

Without loss of generality, we consider an incident electromagnetic wave  in the $z$-direction with the vector potential $\vec A(t)=(A_x \cos(\omega t),A_y\sin(\omega t+\phi),0)$. Following standard routes~\cite{OkaAoki}, the Peierls substitution  $H_{\rm lin} ( \vec q ) \rightarrow H_{\rm lin} ( \vec q+e\vec A(t)) $ and averaging over one drive cycle leads to the effective Hamiltonian \cite{supple}:
\begin{align}
H_{\rm eff} (\vec q) =
\begin{pmatrix}
H_{\rm WF} (\vec q) & O(q)+O(A^2)\\
O(q)+O(A^2)&D_0+O(q)+O(A^2)
\end{pmatrix},
\end{align}
where the $2\times2$ Weyl-Floquet part reads
\ba
H_{\rm WF} (\vec q) &=&H_{\rm W} (\vec q ) -\frac{e^2A_xA_y\cos\phi}{\omega}\times \nn
&& \Big[\epsilon_{ijk}\beta_{xi}\beta_{yj}\sigma_k -\frac{i}{2}(C_xC_y^{\dagger}-C_yC_x^{\dagger})\Big].
\label{eq_Weyl_Floquet}
\ea
Higher-order terms $O(q^2)+O(qA^2)+O(A^4)$ are ignored. Clearly, the photoinduced effect is maximized for a CP light when $\cos \phi = \pm 1$, and vanishes for a linearly polarized beam.

The CP light has two contributions to the Weyl-Floquet Hamiltonian in Eq.~(\ref{eq_Weyl_Floquet}). The first ($\beta^2$) term originates entirely from the two-band Weyl Hamiltonian $H_{\rm W} (\vec q)$, while the second term proportional to $C_x C_y^\dagger - C_y C_x^\dagger$ is due to high-energy band-mixing and must be kept in the construction of a general Weyl-Floquet effective model. For later convenience, we re-parameterize $-(i/2) (C_xC_y^{\dagger}-C_yC_x^{\dagger}) =\kappa_0\sigma_0+\kappa_i\sigma_i$ with four real numbers $\kappa_i = (-i /4) \mbox{Tr}[\sigma_i(C_xC_y^{\dagger}-C_yC_x^{\dagger})]$. Therefore, the influence of the drive to $A^2$-order on the $2\times2$ Weyl band can be summarized as:
\ba
H_{\rm WF}(\vec q)&=&(q_i-\delta q_i)\alpha_i \sigma_0+(q_i-\delta q_i)\beta_{ij}\sigma_j-\delta\mu\cdot\sigma_0 , \nn
\delta q_i&=&\frac{e^2 A_xA_y\cos\phi}{\omega}[\epsilon_{jkl}\beta_{xj}\beta_{yk}+\kappa_l](\beta^{-1})_{li}, \nn
\delta\mu&=&-\delta \vec q\cdot \vec \alpha+\frac{e^2A_xA_y\cos\phi}{\omega}\kappa_0. \label{eq:q-and-mu-shifts}
\ea
$\delta q_i$ and $\delta \mu$ represent the shifts of Weyl momenta and chemical potential, respectively.

The photoinduced Weyl node shift has an immediate consequence on the electronic transport. According to Ref.~\cite{Yang:2011p75129}, for a Weyl system where the chemical potential situates at the nodes (i.e. $\mu=0$), the leading order change of the anomalous Hall conductivity is governed by the momentum shifts of every Weyl node as:
\begin{align}
\delta \sigma_{ij}=\frac{e^2}{2\pi h}\epsilon_{ijk}\delta \nu_k,\ \mbox{      where }
\delta \nu_k=\sum_{I}\chi_W^{(I)}\cdot \delta q^{(I)}_k .\label{eq:anomalous_Hall}
\end{align}
Different Weyl nodes are labeled by the superscript $I$ and $\delta \vec \nu$ is the change of the Chern vector \cite{Yang:2011p75129}. Provided that the momentum shifts $\delta q^{(I)}_k$ among Weyl nodes of different chiralities $\chi_W^{(I)}$ do not cancel out, one would expect the AHE induced at the $A^2$-order proportional to the intensity of the incident beam. Note that the photoinduced self-doping ($\delta\mu \neq 0$) of the Weyl nodes contributes to $\delta \sigma_{ij}$ at a higher order $O(\delta\mu^2)=O(A^4)$~\cite{Yang:2011p75129} of the drive intensity, which is negligible. If $\vec \kappa^{(I)}=0$, one can show the $z$-component of the Chern vector shift to be
\ba
\delta\nu_z=\frac{e^2A_xA_y\cos\phi}{\omega}\sum_{I, i}\frac{[\mbox{Cof}(\beta^{(I)})_{zi}]^2}{|\det(\beta^{(I)})|},
\ea
where $\mbox{Cof}(\beta^{(I)})_{ij}$ is the $ij$ element of the cofactor matrix of $\beta^{(I)}$. The alternating signs of $\chi^{(I)}$ between a node and an anti-node, or a monopole and an anti-monopole, are precisely cancelled by the same sign change in $\delta q^{(I)}_z$. As such, each Weyl node has a positive-definite contribution to $\delta\nu_z$. Consequently, a finite Hall conductivity is induced in the plane perpendicular to the incident beam and its sign is determined by the photon chirality through $\sigma_{xy} \propto \cos \phi$.

In practice, the pre-drive chemical potential $\mu$ may not cross the Weyl node as we discussed so far. As long as $\mu$ before the drive lies in the linearly dispersive regime near the node, one can show that the our resultant $\delta \sigma_{ij}$ is unaffected \cite{supple}. In fact, using a frequency of excitation $0<\omega<2\mu$ for slightly electron-doped Weyl node would ensure no absorption, as assumed here, and thus should lead to a more straightforward analysis of the experimental results regarding the photoinduced AHE.
\\

%%%%%%%%%%%%%%%%%%%%%%%%%%%%%%%%%%%%%%%%%%%%%%%%%%%%%%%%%%%%%%%%%%%%%%%%%%%%%%%%%%%%%%%%%%%%%%%%%%%%%%%%%%%%%%%%%%%%%%%%%%%%%%%%%%%%%%%%%%
{\it Symmetry consideration for multiple Weyl nodes}: In a typical Weyl material, there are multiple sets of Weyl nodes related by the space group and time-reversal (TR) symmetries~\cite{wan11,Dai:2015p11029}. The parameters $\vec \alpha$, $\beta$, and $C_i$ characterizing symmetry-related nodes transform accordingly. We now derive their transformation rules and discuss how the nodal shifts $\delta \vec q$ and $\delta\mu$ among the symmetry-related nodes are correlated.

A pair of Weyl points $\vec k_{\rm W}^{(I)}$ (with $I=1,2$) related by some space-group symmetry satisfy $R \vec k_{W}^{(1)}=\vec k_{W}^{(2)}$, where $R\in O(3)$ defines the particular symmetry operation. The parameters quantifying the Weyl nodes transform according to
\ba
\alpha^{(2)}_i=R_{ij}\alpha^{(1)}_j, ~ \beta^{(2)}_{ij}= R_{ik}\beta^{(1)}_{kj}, ~C^{(2)}_{i}= R_{ij}C^{(1)}_{j}.  \label{eq:unitary_symm_relations}
\ea
As a result, the chiralities of nodes related by a rotation ($\mbox{Det}[R] =+1$) are the same, whereas a mirror or inversion operation ($\mbox{Det}[R] =-1$) produces a sign difference. On the other hand, nodes related by an anti-unitary symmetry involving TR operation satisfy
\ba
\alpha^{(2)}_i \!=\!R_{ij}\alpha^{(1)}_j,\ \beta^{(2)}_{ij}\!=\!R_{ik}\beta^{(1)}_{kl}\xi_{lj},\ C^{(2)}_{i}\!=\! R_{ij}[ C^{(1)}_j ]^* .
\label{eq:antiunitary_symm_relations}
\ea
Due to the presence of $\xi=(\xi_{lj}) = \mbox{diag}(1,-1,1)$, the chiralities of these two nodes are related by $-\mbox{Det}[R]$. For instance, the TR operation with $R={\rm diag} (-1,-1,-1)$ gives the same chirality for the nodes at $\vec k_{\rm W}$ and $R\vec k_{\rm W} = -\vec k_{\rm W}$.

Some proposed Weyl semimetals break TR symmetry but preserve inversion (I). Our symmetry analysis predicts $\vec\alpha^{(2)}=-\vec\alpha^{(1)}$, $\beta^{(2)}=-\beta^{(1)}$, $\kappa_0^{(2)} =\kappa_0^{(1)}$ and $\vec \kappa^{(2)}=\vec \kappa^{(1)}$ for a pair of I-related nodes and thus, using Eq.~(\ref{eq:q-and-mu-shifts}),
\ba
\delta \vec q^{(2)}= -\delta \vec q^{(1)}, ~ \delta \mu^{(2)}= \delta \mu^{(1)}. ~ ({\rm I\!-\!related})
\ea
Since these two nodes must have opposite chiralities ($\mbox{Det}[R]=-1$), the product $\chi^{(I)} \delta \vec q^{(I)}$ is identical for both of them. Hence, from Eq.~(\ref{eq:anomalous_Hall}), these two nodes contribute additively to the Hall conductivity. A similar conclusion holds for an I symmetry breaking Weyl materials for which TR is a good symmetry. In this case we find $\vec\alpha^{(2)}=-\vec\alpha^{(1)}$, $\beta^{(2)}=-\beta^{(1)}\cdot\xi$, $\kappa_0^{(2)} =-\kappa_0^{(1)}$ and $\vec \kappa^{(2)}=-\xi\cdot\vec \kappa^{(1)}$ for TR-related nodes, and thus
\ba
\delta \vec q^{(2)}= \delta \vec q^{(1)}, ~ \delta \mu^{(2)}= -\delta \mu^{(1)}. ~({\rm TR\!-\!related})
\ea
Since $\chi^{(1)}=\chi^{(2)}$, Eq.~(\ref{eq:anomalous_Hall}) once again predicts a constructive contributions from the TR-related nodes.
\\

\begin{figure}[t]
\begin{center}
\includegraphics[angle=0, width=1\columnwidth]{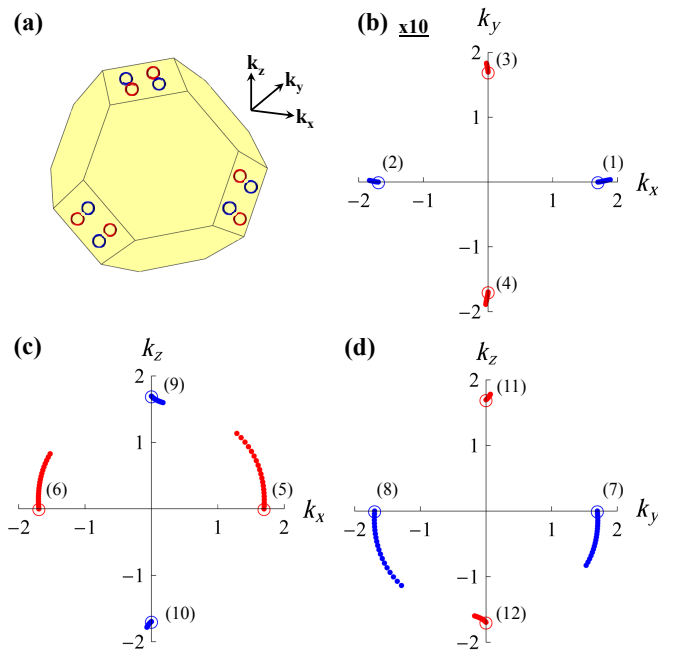}
\caption{(Color online) (a) The 12 Weyl nodes in the first Brillouin zone of the undriven lattice Weyl model studied in Ref.~\cite{Ojanen:2013p245112}. (b), (c), and (d) Photoinduced Weyl node shifts in the presence of a CP drive along the $z$-direction. Nodes with positive (negative) chirality are represented by red (blue) dots, and (I) labels the I-th Weyl node (see Eq.~(\ref{eq:12-weyl-nodes})). Open circles denote the undriven Weyl node positions.  Only projections on the $k_z=2\pi$, $k_y=2\pi$, and $k_x=2 \pi$ planes are shown as we increase $A^2$ from 0 to 0.3. The shifts in (b) are magnified 10 times for a better visualization. Details of the model calculation can be found in \cite{supple}. Parameters used: $(t,\epsilon,\lambda,\omega)=(1,3,1,1.1)$.  }
\label{fig_node_shift}
\end{center}
\end{figure}

%%%%%%%%%%%%%%%%%%%%%%%%%%%%%%%%%%%%%%%%%%%%%%%%%%%%%%%%%%%%%%%%%%%%%%%%%%%%%%%%%%%%%%%%%%%%%%%%%%%%%%%%%%%%%%%%%%%%%%%%%%%%%%%%%%%%%%%%%%
{\it Lattice model analysis}: To illustrate our findings, we now proceed to study a concrete lattice model for Weyl semimetal of inversion symmetry breaking type~\cite{Ojanen:2013p245112}. The model is parameterized by the hopping (t), spin-orbit coupling ($\lambda$) and inversion breaking potential ($\epsilon$) \cite{supple}. The bare system possesses $12$ Weyl nodes labeled as (see Fig.~\ref{fig_node_shift}(a)):
\begin{align}
\vec k_{W-}^{(1,2)}&=(\pm k_0,0,2\pi),& \vec k_{W+}^{(3,4)}&=(0, \pm k_0,2\pi), \notag\\
\vec k_{W+}^{(5,6)}&=(\pm k_0,2\pi,0), & \vec k_{W-}^{(7,8)}&=(2\pi, \pm k_0,0),  \notag\\
\vec k_{W-}^{(9,10)}&=(0,2\pi, \pm k_0),& \vec k_{W+}^{(11,12)}&=(2\pi,0, \pm k_0),
\label{eq:12-weyl-nodes}
\end{align}
along with their chiralities as subscripts $\pm$ and a characteristic momentum $k_0=2\sin^{-1} [\epsilon/(4\lambda)]$. In the presence of a CP drive, each node undergoes a momentum shift given by:
\ba
-\delta\vec q_-^{(1,2)}=\delta\vec q_+^{(3,4)}& = & \frac{e^2A^2}{\omega}\frac{8\eta(1-\eta)\lambda^2}{\epsilon} \hat{z} , \notag\\
\delta\vec q_+^{(5,6)}=-\delta\vec q_-^{(7,8)} & = & \frac{e^2A^2}{\omega}\frac{8\eta(1+\eta)\lambda^2}{\epsilon} \hat{z} ,  \notag\\
-\delta\vec q_-^{(9,10)} = \delta\vec q_+^{(11,12)} & = & \frac{e^2A^2}{\omega}\frac{\epsilon t^2}{16\eta \lambda^2} \hat{z},
\ea
with $\eta = \sqrt{1- \epsilon^2/ 16\lambda^2 }$. The momentum shifts clearly correlate with the chiralities. The photoinduced self-dopings are:
\begin{align}
-\delta\mu^{(9)}=\delta\mu^{(10)}=\delta\mu^{(11)}=-\delta\mu^{(12)}=\frac{e^2A^2}{\omega}\frac{\epsilon^2}{16},
\end{align}
and $\delta \mu^{(1-8)}=0$. The anomalous Hall conductivity obtained is (up to $A^2$-order):
\begin{align}
\delta\sigma_{xy}&=\frac{e^2}{2\pi h} \frac{e^2A^2}{\omega}\left(\frac{64\eta\lambda^2}{\epsilon}+\frac{\epsilon t^2}{4\eta\lambda^2}\right),
\end{align}
which agrees with the expectation of the low-energy analysis.

\begin{figure}[t]
\begin{center}
\includegraphics[angle=0, width=1\columnwidth]{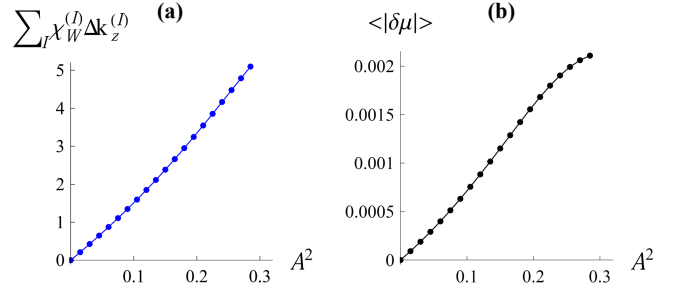}
\caption{(Color online) The field dependence for (a) the total $\Delta k_z$ weighted by chiralities  and (b) the absolute self-doping $|\delta \mu|$ averaged over 12 Weyl nodes for the driven Weyl lattice model. Same parameters used as in Fig.~\ref{fig_node_shift}.}
\label{fig_Hall_vector}
\end{center}
\end{figure}

So far all our discussion have been done within the effective Hamiltonian scheme valid for a small driving amplitude. A more stringent check free from this assumption is offered by solving the full Floquet Hamiltonian whose matrix elements are $\langle n'| H_F |n\rangle =  H_{n-n'}  + n\omega \delta_{n,n'} I_4 $ with $n(n')$ denoting the $n(n')$-th Floquet band. The topological character of the Weyl nodes indeed protects them against the drive, and only their locations are shifted continuously in a chirality-dependent way. Figure~\ref{fig_node_shift}(b-d) demonstrate the Weyl node shifts obtained from diagonalizing the full Floquet Hamiltonian as we increase the driving field strength. It is obvious that nodes with opposite chirality have opposite shifts in $k_z$, leading to an imbalance of $\delta \nu_z$ and eventually, a net anomalous Hall conductivity. The overall field dependence of the sum of $\Delta k_z$ weighted by chiralities is shown in Fig.~\ref{fig_Hall_vector}(a), which exhibits the anticipated $A^2$ increment (we have put $e=1$). This quantity equals to the Chern vector $\delta \nu_z$ in the small $A$ regime (Eq.~(\ref{eq:anomalous_Hall})). On the other hand, the resultant self-doping shows the expected $A^2$ dependence as illustrated in Figure~\ref{fig_Hall_vector}(b). However, it only corresponds to a higher order effect and does not affect the photoinduced AHE.
\\

%%%%%%%%%%%%%%%%%%%%%%%%%%%%%%%%%%%%%%%%%%%%%%%%%%%%%%%%%%%%%%%%%%%%%%%%%%%%%%%%%%%%%%%%%%%%%%%%%%%%%%%%%%%%%%%%%%%%%%%%%%%%%%%%%%%%%%%%%%
{\it Application to TaAs family of Weyl semimetals}: TaAs was recently shown to host 24 Weyl nodes in the Brillouin zone~\cite{Dai:2015p11029,Lv:2015kp,Huang.S.M:2015}. There are two nonequivalent sets of Weyl points~\cite{Dai:2015p11029} that we denote as: the P-set consisting of $8$ nodes at $(\pm k^P_x, \pm k^P_y,0)$ and $(\pm k^P_y, \pm k^P_x,0)$, and the Q-set with $16$ nodes at $(\pm k^Q_x, \pm k^Q_y, \pm k_z^Q)$ and $(\pm k^Q_y, \pm k^Q_x, \pm k_z^Q)$. The values of $k^P_x, k^P_y, k^Q_x,k^Q_y,k_z^Q$ for various TaAs family of materials can be found in Ref.~\cite{Dai:2015p11029}. Applying the symmetry transformation rules derived before, we find a simple rule for the node shifts for both Weyl sets:
\ba
\delta q_{z}^{(P,I)} &=& \chi^{(P,I)} \delta q_z^P  , ~~ 1 \le I \le 8, \nn
\delta q_z^{(Q,I)} &=& \chi^{(Q,I)} \delta q_z^Q  , ~~ 1 \le I \le 16,
\ea
where $\delta q_z^{P(Q)} \propto A^2$ is the same for all $8(16)$ symmetry-related nodes. The $z$-direction momentum shift faithfully follows the chirality $\chi^{(I)}$ of each Weyl node, whereas the same is not true for the shifts in the xy-plane. Therefore, the overall photoinduced AHE for TaAs family is:
\begin{align}
\sigma_{xy}&=\frac{e^2}{2\pi h} \left( 8 \times \delta q^P+16\times \delta q^Q \right) .
\end{align}

%%%%%%%%%%%%%%%%%%%%%%%%%%%%%%%%%%%%%%%%%%%%%%%%%%%%%%%%%%%%%%%%%%%%%%%%%%%%%%%%%%%%%%%%%%%%%%%%%%%%%%%%%%%%%%%%%%%%%%%%%%%%%%%%%%%%%%%%%%
{\it Experimental Realization:} To guide future experimental efforts to observe these effects, we discuss the conditions needed and likelihood of success. In the supplemental we detail our calculations of the magnitude of the Hall voltage \cite{supple}, where we assume the use of a CW CO$_{2}$ laser, with a roughly constant illumination across the contacts, spaced $\approx 100~\mu$m apart. Using established values for the Fermi velocity~\cite{Yang:2015ev,HongDing,2015NatPh..11..645S}, reflectance, optical~\cite{2015arXiv151000470X} and DC conductivities~\cite{Huang:2015gy}, we find a $100~\rm nm$ thick film~\cite{thickfilm} of TaAs would produce a Hall signal $\approx 130~nV$ at room temperature, with a DC current of 1A and 1W laser power. We note that the small penetration depth requires the use of thinner samples to ensure most of the current is modulated by the light (the induced voltage is inversely proportional to the square of the thickness).  While such signals should be straight forward to detect, even higher values may be possible through the use of pulsed lasers with higher peak electric fields, with the Hall signal detected through Faraday rotation measurements.
\\

%%%%%%%%%%%%%%%%%%%%%%%%%%%%%%%%%%%%%%%%%%%%%%%%%%%%%%%%%%%%%%%%%%%%%%%%%%%%%%%%%%%%%%%%%%%%%%%%%%%%%%%%%%%%%%%%%%%%%%%%%%%%%%%%%%%%%%%%%%
{\it Conclusion}: An assortment of analysis based on Floquet theory is carried out on models of Weyl semimetals to argue that the AHE is induced generically by applying an AC electromagnetic field of a definite chirality, in the plane orthogonal to the incident beam. This photoinduced AHE originates from the remarkable monopole nature of Weyl nodes. The induced Hall conductivity scales linearly and continuously with the field intensity. The nodal shift itself may be observable by the pump-probe ARPES, while the photoinduced AHE can be detected via DC transport or Faraday rotation experiment on films. Conceivably, the AC-field-driven AHE may exist in other materials possessing non-trivial band topology such as ferromagnetic metals~\cite{RevModPhys.82.1539}.

{\it Acknowledgements} -  PAL acknowledges support from DOE Grant No. DE-FG02-03- ER46076. KSB acknowledges support from the National Science Foundation under Grant No. DMR-1410846. JHH is supported by the NRF grant (No. 2013R1A2A1A01006430). YR is supported by the Alfred P. Sloan fellowship and National Science Foundation under Grant No. DMR-1151440.

\bibliography{Weyl_Floquet}

\begin{thebibliography}{31}
\bibitem [{thi()}]{thickfilm}%
  \BibitemOpen
  \href@noop {} {}\bibinfo {note} {The proposed film has a thickness larger
  than the lattice spacing by two orders of magnitude so that the momentum
  along the width is still a good quantum number.}\BibitemShut {Stop}%
\end{thebibliography}

%%%%%%%%%%%%%%%%%%%%%%%%%%%%%%%%%%%%%%%%%%%%%%%%%%%%%%%%%%%%%%%%%%%%%%%%%%%%%%%%%%%%%%%%%%%%%%%%%%%%%%%%%%%%%%%%%%%%%%%%%%%%%%%%%%%%%%%%%%%%%%
\clearpage
\begin{widetext}

\begin{center}
\large{\bf Supplemental Material:\\ When Chiral Photons Meet Chiral Fermions-- \\
Photoinduced Anomalous Hall Effects in Weyl Semimetals}\\
\vspace{14pt}
\normalsize{Ching-Kit Chan, Patrick A. Lee, Kenneth S. Burch, Jung Hoon Han, and Ying Ran}
\end{center}
%\vspace{14pt}

%%%%%%%%%%%%%%%%%%%%%%%%%%%%%%%%%%%%%%%%%%%%%%%%%%%%%%%%%%%%%%%%%%%%%%%%%%%%%%%%%%%%%%%%%%%%%%%%%%%%%%%%%%%%%%%%%%%%%
\section{Effective Weyl-Floquet Hamiltonian}\label{sec:tight_binding}

Here, we provide more details about the derivation of the effective Weyl-Floquet Hamiltonian in the main text. We start with a general $N$-band tight-binding model $H(\vec k)$ of a Weyl semimetal (SM). For convenience, we choose the energy eigenstates at $\vec k_W$ as the basis and shift the energy so that the Weyl node is at zero energy. The linearized Hamiltonian $H_{\text{lin}}$ in this basis has the follow block form:
\begin{align}
U^{\dagger} H(\vec q+\vec k_W) U
=H_{\text{lin}}+O(q^2)\notag
=\begin{pmatrix}
q_i\alpha_i \sigma_0+q_i\beta_{ij}\sigma_j & q_i C_{i}\\
q_i C^{\dagger}_{i}& D_0+q_i D_i
\end{pmatrix}+O(q^2),
\end{align}
where the columns of $U$ are the eigenstates of $H(\vec k_W)$ with the first two columns corresponding to the two Weyl bands. We introduce a real 3-vector $\vec \alpha$ and a $3\times 3$ real matrix $\beta$ to parameterize the Weyl band structure. Here, the Hermitian matrices $D_i$ correspond to the $(N-2)$ high energy bands and $D_0$ is diagonal, while $C_i$ are $2\times (N-2)$ matrices describing the mixing between the Weyl and high energy bands.

Consider a electromagnetic field propagating along the $\hat z$-direction. After coupling $H_{\text{lin}}$ with the gauge field: $\vec q\rightarrow \vec q+e\vec A(t)$, with $\vec A(t)=(A_x \cos(\omega t),A_y\sin(\omega t+\phi),0)$, we obtain:
\begin{align}
 H_{\text{lin}} \rightarrow H_{\text{lin}}+H_{-1}e^{i\omega t}+H_{+1}e^{-i\omega t},
\end{align}
with $H_{-1}=H_{+1}^{\dagger}$ and
\begin{align}
 H_{-1}=\frac{e}{2}
 \begin{pmatrix}
(A_x\alpha_x-ie^{i\phi}A_y\alpha_y)\sigma_0+A_x\beta_{x j}\sigma_j-ie^{i\phi}A_y \beta_{y j}\sigma_j& A_x C_x-i e^{i\phi} A_yC_y\\
A_x C_x^{\dagger}-i e^{i\phi} A_y C_y^{\dagger}& A_x D_x-i e^{i\phi}A_y D_y
 \end{pmatrix}
\end{align}
In the perturbative regime $A/\omega \ll 1$, the time-independent effective Hamiltonian near $\vec k_W$ is given by \cite{oka09_s}:
\begin{align}
 H_{\text{eff}}&=H_{\text{lin}}+\frac{1}{\omega}[H_{-1},H_{+1}] \notag \\
 &=  \begin{pmatrix}
  H_{\text{WF}}(\vec q)&O(q)+O(A^2)\\
  O(q)+O(A^2)&D_0+O(q)+O(A^2)
 \end{pmatrix},
\end{align}
with
\begin{align}
 H_{\text{WF}}(\vec q)= & q_i\alpha_i \sigma_0+q_i\beta_{ij}\sigma_j -\frac{e^2A_xA_y\cos\phi}{\omega}\big[\epsilon_{ijk}\beta_{xi}\beta_{yj}\sigma_k -\frac{i}{2}(C_xC_y^{\dagger}-C_yC_x^{\dagger})\big],
\end{align}
which is the result in the main text. Perturbation theory tells that $H_{\text{WF}}(\vec q)$ is the correct 2-band effective Weyl-Floquet Hamiltonian up to $O(q^2)+O(qA^2)+O(A^4)$. Note that the $\beta$ term inside the bracket can be understood from a 2-band $H_{\text{W}}(\vec q)$ itself, while the $C_i$ terms are contributed by the linear-$q$ mixing between the Weyl and high energy bands.

%%%%%%%%%%%%%%%%%%%%%%%%%%%%%%%%%%%%%%%%%%%%%%%%%%%%%%%%%%%%%%%%%%%%%%%%%%%%%%%%%%%%%%%%%%%%%%%%%%%%%%%%%%%%%%%%%%%%%
\section{Calculations for the driven lattice model}

In this section, we detail the calculation for the Weyl model studied in the main text. We adopt the four-band model proposed by Ojanen \cite{ojanen13_s} on a diamond lattice with an inversion symmetry breaking field. In momentum space, the undriven Hamiltonian is given by:
\begin{eqnarray}
H_0=\sum_{i=1}^5 D_i(\vec k) \Gamma_i - i \epsilon \Gamma_{4} \Gamma_{5},
\label{eq_ojanen_model}
\end{eqnarray}
where $\Gamma_i$ are Gamma matrices acting on the spin and orbital degrees of freedom. We take $\Gamma_1 = \sigma_z \otimes \tau_x $, $\Gamma_2 = \sigma_z \otimes \tau_y$, $\Gamma_3 = \sigma_z \otimes \tau_z$, $\Gamma_4 = \sigma_x$ and $\Gamma_5 = \sigma_y$. Here,
\begin{eqnarray}
D_{1}(\vec k)&=&2 \lambda \sin\frac{k_x}{2} \left(\cos\frac{k_z}{2}-\cos\frac{k_y}{2}\right), \notag \\
D_{2}(\vec k)&=&2 \lambda \sin\frac{k_y}{2} \left(\cos\frac{k_x}{2}-\cos\frac{k_z}{2}\right), \notag \\
D_{3}(\vec k)&=&2 \lambda \sin\frac{k_z}{2} \left(\cos\frac{k_y}{2}-\cos\frac{k_x}{2}\right), \notag \\
D_{4}(\vec k)&=&t  \left(1+\cos\frac{k_x+k_y}{2}+\cos\frac{k_x+k_z}{2}+\cos\frac{k_y+k_z}{2}\right), \notag \\
D_{5}(\vec k)&=&t  \left(\sin\frac{k_x+k_y}{2}+\sin\frac{k_x+k_z}{2}+\sin\frac{k_y+k_z}{2}\right),
\end{eqnarray}
where $t$, $\lambda$, and $\epsilon$ parameterize the nearest-neighbor hopping, spin-orbit couplings, and the inversion breaking potential, respectively. When $|\epsilon| < 4 |\lambda|$, this system possesses 6 pairs of Weyl nodes \cite{ojanen13_s} at $\vec k_W =(\pm k_0,0,2\pi)$, $(\pm k_0,2\pi,0)$, $(0,\pm k_0,2\pi)$, $(2\pi,\pm k_0,0)$, $(2\pi,0,\pm k_0)$ and $(0,2\pi, \pm k_0)$, with $k_0= 2 \sin^{-1}[\epsilon/(4\lambda)]$ (see Fig.~2(a) in the main text). However, due to the TR symmetry, the undriven system does not have a net anomalous Hall conductivity (i.e. $\vec \nu= \sum_I \chi^{(I)}_W k^{(I)}_W= 0 $).

In the presence of a circularly polarized light, the Hamiltonian becomes time-dependent through the Peierls substitution $\vec k \rightarrow \vec k + A (\cos\omega t, \sin \omega t, 0)$. This allows the definition of the Floquet Hamiltonian \cite{shirley65_s}: $ H_F \left|\phi_n(t)\right\rangle = \left[H(t)-i\partial_t \right] \left|\phi_n(t)\right\rangle = \epsilon^F_{n} \left|\phi_n(t)\right\rangle $, where $\epsilon^F_{n}$ and $\left|\phi_n(t)\right\rangle$ are the Floquet eigenenergies and eigenvectors satisfying the time-periodicity $\left|\phi_n(t)\right\rangle = \left|\phi_n(t+2 \pi/\omega)\right\rangle$. Fourier expanding $\left|\phi_n(t)\right\rangle$ further enables us to express the Floquet Hamiltonian as $\langle n'| H_F |n\rangle =  H_{n-n'}  + n\omega \delta_{n,n'} I_4 $, where $\left| n \right\rangle$ denotes the eigenmode with frequency $n \omega$.

The Fourier expansion of $H(t) =\sum_n H_{-n} e^{i n\omega t}$ are analytically tractable. The result is:
\begin{eqnarray}
H_{-n}=\sum_{i=1}^5 d_{i,-n}(\vec k) \Gamma_i - i \epsilon \Gamma_{4} \Gamma_{5} \delta_{n,0},
\label{eq_drivenojanen_model}
\end{eqnarray}
where
\begin{eqnarray}
d_{1,-n}(\vec k)&=& \lambda \left\{ 2 J_n\left(\frac{A}{2}\right) \cos\frac{k_z}{2} \sin\left(\frac{k_x+n\pi}{2}\right)- J_n\left(\frac{A}{\sqrt 2}\right) \left[e^{-\frac{i n \pi}{4} } \sin\left(\frac{k_x+k_y+ n\pi}{2}\right) + e^{\frac{i n \pi}{4} } \sin\left(\frac{k_x-k_y+ n\pi}{2}\right)  \right] \right\}, \notag \\
d_{2,-n}(\vec k)&=& \lambda \left\{ - \frac{2}{i^n} J_n\left(\frac{A}{2}\right) \cos\frac{k_z}{2} \sin\left(\frac{k_y+n\pi}{2}\right)+ J_n\left(\frac{A}{\sqrt 2}\right) \left[e^{-\frac{i n \pi}{4} } \sin\left(\frac{k_x+k_y+ n\pi}{2}\right) - e^{\frac{i n \pi}{4} } \sin\left(\frac{k_x-k_y+ n\pi}{2}\right)  \right] \right\}, \notag \\
d_{3,-n}(\vec k)&=& 2 \lambda J_n\left(\frac{A}{2}\right) \sin\frac{k_z}{2} \left[(-i)^n \cos\left(\frac{k_y+ n\pi}{2}\right) - \cos\left(\frac{k_x+ n\pi}{2}\right)  \right] , \notag \\
d_{4,-n}(\vec k)&=& t\left\{\delta_{n,0} + J_n\left(\frac{A}{\sqrt 2}\right) e^{-\frac{in\pi}{4}} \cos\left(\frac{k_x+k_y+ n\pi}{2}\right) + J_n\left(\frac{A}{2}\right) \left[ \cos\left(\frac{k_x+k_z+ n\pi}{2}\right) + (-i)^n \cos\left(\frac{k_y+k_z+ n\pi}{2}\right)  \right] \right\}, \notag \\
d_{5,-n}(\vec k)&=& t\left\{J_n\left(\frac{A}{\sqrt 2}\right) e^{-\frac{i n \pi}{4}} \sin\left(\frac{k_x+k_y+ n\pi}{2}\right) + J_n\left(\frac{A}{2}\right) \left[ \sin\left(\frac{k_x+k_z+ n\pi}{2}\right) + (-i)^n \sin\left(\frac{k_y+k_z+ n\pi}{2}\right)  \right] \right\}.
\end{eqnarray}
The Weyl node shifts in the main text (Fig.~2 and ~3) are obtained by numerically solving the full Floquet Hamiltonian. For demonstration purposes, we used the parameter values $t=1$, $\epsilon=3$, $\lambda=1$, $\omega=1.1$ and $A$ up to $\sqrt{0.3}$. Our findings are qualitatively unchanged in different parameter regimes, as long as the bare system possesses Weyl nodes (i.e. $|\epsilon| < 4 |\lambda|$) and the driving amplitude is not large enough to cause topological phase transitions by merging Weyl nodes. In our numerical calculations using $A$ up to $\sqrt {0.3}$, the Floquet spectrum converges rapidly by including up to $n=\pm 2$ Floquet eigenmodes.

%%%%%%%%%%%%%%%%%%%%%%%%%%%%%%%%%%%%%%%%%%%%%%%%%%%%%%%%%%%%%%%%%%%%%%%%%%%%%%%%%%%%%%%%%%%%%%%%%%%%%%%%%%%%%%%%%%%%%
\section{Realistic Experimental Estimations}

We provide more supports and details for the estimated transport measurement discussed in the main text. First, let us define the spacing between the voltage contacts as $l_{y}$, current contacts as $l_{x}$, the sample thickness $t$ and penetration depth $\delta$. We assume the light equally illuminates the contact region with a power P. The sample has a refraction index $n(\omega)$, a reflectance $R(\omega)$, a real part of the optical conductivity $\sigma_{1}(\omega)$ and a dc longitudinal conductivity $\sigma_{xx}$. Furthermore we consider an experiment using a CO$_{2}$ CW laser ($\hbar\omega =117~\rm meV$) and a constant current I$_{x}$. The measured Hall voltage is:
\begin{equation}
\label{eq:voltage}
V_{y} \approx \frac{\sigma_{xy} \delta /t}{\sigma_{xx}^{2}+ (\sigma_{xy} \delta/t)^{2}} \times  \frac{l_{y}}{l_{x}t}  \times I_{x}
\end{equation}
where for simplicity we assume a uniform $\sigma_{xy}$ throughout the penetration depth, while the current is evenly spread through the entire thickness (hence the factor $\delta / t$ in $\sigma_{xy}$). Since $\sigma_{xx} \gg \sigma_{xy}$, we only need to be concerned with $\sigma_{xy}\delta/(\sigma_{xx}^{2}t)$. Each Weyl node contributes a $\sigma_{xy}=\frac{e^2}{2\pi h}\frac{ \hbar v_{F} e^{2}E^{2}}{(\hbar\omega)^{3}}$, and we note the relations $ E^2=\frac{2P(1-R)}{l_{x}l_{y} n(\omega)\epsilon_{0}c}$ and $\delta(\omega)=\frac{n(\omega)\epsilon_{0}c}{\sigma_{1}(\omega)}$. Thus, accounting for the $24$ Weyl nodes in TaAs, we find:
\begin{equation}
\label{eq:voltage3}
V_{y}\approx  \frac{ 24}{ \sigma_{xx}^2 t^2 } \times \frac{e^2}{2\pi h} \times \frac{\hbar v_{F} e^2}{(\hbar\omega)^{3}} \times \frac{2 P(1-R)}{l_{x}l_{y}\sigma_{1}(\omega)}  \times  \frac{l_{y}}{l_{x}}  \times  I_{x}
\end{equation}
From previous measurements of the optical \cite{2015arXiv151000470X_s} and DC conductivities \cite{Huang:2015gy_s} and $v_{F}$ of TaAs \cite{Yang:2015ev_s,HongDing_s,2015NatPh..11..645S_s}, we take the values at $300~\text{K}$: $R\approx 0.8$, $\sigma_1\approx 2\times 10^5~\Omega^{-1} m^{-1}$, $\sigma_{xx}\approx 1.7\times 10^6~\Omega^{-1} m^{-1}$ and an average $\hbar v_F\approx 2~\text{eV} \AA $. Considering a sample with $t=100~\text{nm}$, $l_x=l_y=100~\mu\text{m}$ and using P=1~W, $I_{x}$=1~A, the resultant Hall voltage is $V_y\approx 130$~nV, which is easily observable in realistic experiments.

As we remark in the main text, the AHE is also measurable in a pump-probe experiment. Here, we give an order-of-magnitude estimation for $\sigma_{xy}$ in a pulsed laser experiment. Based on the Floquet topological insulator~\cite{wang13_s}, we take a pump laser energy $\hbar \omega \sim 0.12~$eV and an electric field intensity $eA \sim 0.01-0.02~\AA^{-1}$, together with the group velocity for TaAs used above, the resultant $\sigma_{xy} \sim 25-100$ $\Omega^{-1} \rm cm^{-1}$ .

%%%%%%%%%%%%%%%%%%%%%%%%%%%%%%%%%%%%%%%%%%%%%%%%%%%%%%%%%%%%%%%%%%%%%%%%%%%%%%%%%%%%%%%%%%%%%%%%%%%%%%%%%%%%%%%%%%%%%
\section{Doping effects}

Now, we explain the consequences of doping to the photoinduced AHE. For a better comparison, let us recap the findings for $\sigma_{xy}$ in the main text, where the chemical potential lies at the Weyl nodes ($\mu=0$). Before the drive, each Weyl node pair separated by $\vec K$ contributes a $\sigma_{xy}(\mu=A=0)=\frac{e^2}{2\pi h} K_z$. If the system has TR symmetry, the contributions from the TR Weyl node pairs cancel each other and the total $\sigma_{xy}(\mu=A=0)=0$. When there is a circularly polarized drive, each Weyl node pair results in a photoinduced $\delta \sigma_{xy}(\mu=0,A) \sim \frac{e^2}{2\pi h} \frac{2 \hbar v_F e^2 A^2 }{\omega}$ ($v_F$ being the Fermi velocity). The photoinduced contributions from different Weyl pairs always add.

When the initial chemical potential is tuned away from the node, but still within the linearly dispersive regime, our result is unaffected. We first review the argument for the undriven case used in Ref.~\cite{yang11_s}. In the linear regime, the low-energy Hamiltonian of a undriven Weyl node is $H_{\text{eff}} = \sum_i \hbar (\vec v_i \cdot \vec k )\sigma_i - \mu $, where $\vec k$ is the momentum away from the node. The Berry curvature $\vec B(\vec k)$ is odd about the node:
\begin{eqnarray}
\vec B(-\vec k ) = - \vec B(\vec k ),
\label{eq_supple_Weyl_berry}
\end{eqnarray}
as can be shown by performing a formal TR operation: $T H_{\text{eff}}(\vec k) T^{-1} = H_{\text{eff}}(-\vec k)$. As a result, we have
\begin{eqnarray}
\sigma_{xy}(\mu,A=0)- \sigma_{xy}(0,A=0) \sim \frac{e^2}{4\pi^2 h} \int_{0 \leq k \leq k_F} d^3 k \vec B(\vec k) \cdot \hat k_z\sim \frac{e^2}{2\pi h} (c_1 \mu^2) ,
\label{eq_supple_sigmaxy1}
\end{eqnarray}
where $k_F$ is the Fermi momentum and $c_1$ is some microscopic constant. In fact, an ideal Weyl node has $\vec B(\vec k ) \propto \hat k/k^2$ and the above integral vanishes due to the cancellation between contributions from the $+k_z$ and $-k_z$ planes. The $O[\mu^2]$ correction comes from the departure from the ideal Weyl node. Note also that in a TR symmetric system, $\sigma_{xy}(\mu,0)$ is always zero.

We now apply the same argument to the driven system. Since a driven Weyl node is still a monopole, Eq.~(\ref{eq_supple_Weyl_berry}) still holds and we have a similar relation:
\begin{eqnarray}
\sigma_{xy}(\mu,A)- \sigma_{xy}(0,A) \sim \frac{e^2}{2\pi h} (c_1 + c_2 A^2) \mu^2,
\label{eq_supple_sigmaxy2}
\end{eqnarray}
where the $c_2 A^2$ term is a perturbative correction. Combining Eq.~(\ref{eq_supple_sigmaxy1}), (\ref{eq_supple_sigmaxy2}) and $\delta \sigma_{xy}(0,A)  $, we have:
\begin{eqnarray}
\delta \sigma_{xy}(\mu, A) = \sigma_{xy}(\mu,A)- \sigma_{xy}(\mu,0) \sim \frac{e^2}{2\pi h} A^2 \left( \frac{n_W \hbar v_F e^2 }{\omega} + c_2 \mu^2 \right),
\label{eq_supple_sigmaxy3}
\end{eqnarray}
for systems with $n_W$ Weyl nodes. Therefore, as we tune $\mu$ away from the Weyl node, $\delta \sigma_{xy}$ is still proportional to $A^2$ with a $O[\mu^2]$ modification to the prefactor.

When we go beyond the linearly dispersive region, contributions from large momentum enter and the AHE is further modified. Based on the discussion above, we would anticipate a nonlinear change: $c_1  \mu^2 \rightarrow f(\mu,0)$ and $ c_2 A^2 \mu^2 \rightarrow \partial_A f(\mu,A)|_{A=0} A^2 +...$, where $f$ depends on microscopic band-structure details.

To illustrate these effects, we numerically compute $\sigma_{xy}(\mu,A)$ for a two-band model without TR symmetry \cite{deplace12_s}:
\begin{eqnarray}
H(\vec k)=2 t \sin \left(\frac{k_x+k_y}{2}\right) \sigma_x +2 t \sin \left(\frac{k_x-k_y}{2}\right) \sigma_y + \left[2 m_1 - m_2 (\cos k_x +\cos k_y) +2 \cos k_z\right]  \sigma_z.
\label{eq_supple_H2band}
\end{eqnarray}
We consider the parameters $t=m_1=m_2=1$, at which $H$ supports a Weyl node pair at $\vec k = (0,0,\pm \pi/2)$, so that the undriven $\sigma_{xy}(0,0)=\frac{e^2}{2\pi h}\times \pi$. Figure~\ref{fig_supple_doping}(a) and (b) present the Floquet bands and $\sigma_{xy}(\mu,A)$, respectively, based on a perturbative calculation as $\mu$ is tuned away from the node. The results show the expected behavior of $\delta \sigma_{xy}(\mu,A)\propto A^2$ when $\mu$ is small (linearly dispersive regime), and nonlinear corrections for larger $\mu$ and $A$.

We note that, in the calculation above, we have assumed only the bands below the chemical potential are filled. In a general driven system, excited states can also be populated and contribute to $\sigma_{xy}$ (see e.g. \cite{oka09_s}). While this could be an issue for the quantization of the Hall conductance, it does not qualitatively affect our results since we do not require a quantized conductance. The excited state population leads to a correction to the Berry curvature integral (like Eq.~(\ref{eq_supple_sigmaxy1})) and only modifies the proportionality between $\delta \sigma_{xy}$ and $A^2$.

\begin{figure}[t]
\begin{center}
\includegraphics[angle=0, width=0.7\columnwidth]{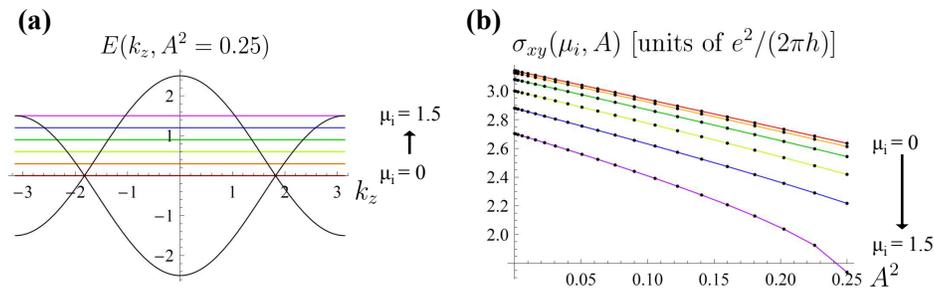}
\caption{Demonstration of the doping effects of the driven two-band Weyl model (see. Eq.~(\ref{eq_supple_H2band})). (a) Weyl-Floquet bands at $k_x=k_y=0$. (b) $\sigma_{xy}(\mu_i,A)$ as the chemical potential $\mu_i$ is increased from $0$ to $1.5$. When $\mu$ is within the linearly dispersive regime, $\delta \sigma_{xy}(\mu,A) \propto A^2$. Parameters used: $t=m_1=m_2=\omega=1$.}
\label{fig_supple_doping}
\end{center}
\end{figure}

\end{widetext}

\end{document}